# Robust Detection of Extremely Thin Lines Using 0.2mm Piano Wire

Jisoo Hong[1], Youngjin Jung[1], Jihwan Bae[1], Seungho Song[1], and Sung-Woo Kang[1]

[1]Department of Industrial Engineering, Inha University, Incheon 22212, Republic of Korea

Corresponding author: Sung-Woo Kang (e-mail: kangsungwoo@inha.ac.kr).

**ABSTRACT** This study developed an algorithm capable of detecting a reference line (a 0.2 mm thick piano wire) to accurately determine the position of an automated installation robot within an elevator shaft. A total of 3,245 images were collected from the experimental tower of H Company, the leading elevator manufacturer in South Korea, and the detection performance was evaluated using four experimental approaches (GCH, GSCH, GECH, FCH). During the initial image processing stage, Gaussian blurring, sharpening filter, embossing filter, and Fourier Transform were applied, followed by Canny Edge Detection and Hough Transform. Notably, the method was developed to accurately extract the reference line by averaging the x-coordinates of the lines detected through the Hough Transform. This approach enabled the detection of the 0.2 mm thick piano wire with high accuracy, even in the presence of noise and other interfering factors (e.g., concrete cracks inside the elevator shaft or safety bars for filming equipment). The experimental results showed that Experiment 4 (FCH), which utilized Fourier Transform in the preprocessing stage, achieved the highest detection rate for the LtoL, LtoR, and RtoL datasets. Experiment 2(GSCH), which applied Gaussian blurring and a sharpening filter, demonstrated superior detection performance on the RtoR dataset. This study proposes a reference line detection algorithm that enables precise position calculation and control of automated robots in elevator shaft installation. Moreover, the developed method shows potential for applicability even in confined working spaces. Future work aims to develop a line detection algorithm equipped with machine learning-based hyperparameter tuning capabilities.

**INDEX TERMS** Line Detection, Piano Wire, Thin Line, Elevator Shaft, Fourier Transform

## I. INTRODUCTION

In modern society, elevators have become an essential mode of transportation in high-rise buildings. With approximately 500,000 units installed globally each year, elevators have established themselves as a critical means of enhancing mobility within buildings[1]. They support the transportation of people and goods across various environments, including residential and public buildings, shopping malls, airports, railway stations, and tourist attractions[2]. As urbanization drives higher population density and reduces the availability of developable real estate, elevators are expected to play an increasingly significant role in shaping the future of cities. Beyond offering convenient vertical transportation, elevators provide critical mobility solutions for the elderly and individuals with disabilities. This highlights the pressing need for safer and more reliable elevator operations.

Advancements in elevator technology have enabled the construction of high-rise buildings by allowing people to easily access upper floors[3]. During the construction of such buildings, elevator installation plays a critical role, but the high-altitude nature of these tasks poses significant risks to workers. The safety and precision of installation work are directly tied not only to the safety of the workers but also to the safety of the users who rely on the elevators after installation. Falls during elevator installation are common and often result in severe injuries or fatalities for workers. To mitigate these accidents, it is essential to either improve the safety of the working environment or adopt robots capable of undertaking these high-risk tasks in place of human workers. Consequently, the development of automated robots for elevator installation has emerged as a crucial solution to enhance worker safety and improve operational efficiency.

Elevator installation work is conducted in various locations, including the elevator shaft, machine room, pit, and landing areas on each floor. As such, fully automating the entire installation process presents significant challenges. Therefore, the primary target for automation is limited to the drilling of anchor holes and the installation of anchor bolts within the elevator shaft, which are critical tasks that consume the most time and pose the highest risk of injury to workers. One of the most crucial aspects of drilling anchor holes in the elevator shaft is ensuring that they are placed in the correct locations. To achieve this, the robot must reliably detect and align with pre-defined reference lines.

Depending on the length of the elevator shaft, the guide rails can extend for several hundred meters[4]. To ensure the safe use of these elevator guide rails, they must undergo periodic measurement. In this context, research has been conducted to measure the straightness of long objects, such as railway tracks, crane rails, and elevator rails. If the robot fails to accurately recognize the reference line, anchor holes may be drilled in incorrect positions. This misalignment can lead to improper installation of the elevator guide rails, reducing the overall accuracy of the installation process. Therefore, the robot's ability to precisely detect and align with the reference line is a critical technical factor that determines the success of the installation work.

In this study, a camera-based approach is employed to recognize reference lines for elevator installation tasks. Specifically, a camera captures images of pre-installed reference lines, and an algorithm is developed to detect these lines within the captured images. This technology offers greater accuracy and consistency compared to traditional manual methods, while also protecting workers from high-risk tasks.

## II. RELATED STUDIES

### A. Study on Installation Techniques in Elevator Shafts

Between 2016 and 2020, a total of 34 workers lost their lives in fall accidents during elevator installation work. Although the total number of elevators in South Korea has exceeded 800,000 units, with over 40,000 new installations annually, the frequency of fatal fall accidents among elevator workers has continued to increase. Although worker negligence accounts for only about 2.5% of elevator-related accidents, the nature of high-altitude work in elevator installation sites presents a relatively high risk of falls. In response to growing concerns over fall accidents raised within the National Assembly, major elevator manufacturers developed a specialized 'elevator installation platform system' aimed at preventing falls, reducing accident rates, and enhancing worker safety.

Additionally, research has been conducted using computer vision techniques and optical sensors to automatically measure the dimensions of elevator shafts without human intervention[5]. Errors in calculating the usable space within the shaft may lead to the production of unusable elevator cabins. Therefore, it is critical to accurately calculate the smallest cross-section within the shaft that determines the maximum floor area of the elevator. This solution is employed to measure the shaft's height and inclination, determine the maximum cabin size that can fit within the inspected shaft, and identify potential deviations caused by bumps or construction defects.

South Korea's elevator industry ranks eighth globally and is the third in terms of new installations. Despite the annual installation of 30,000 to 40,000 new elevators annually, the incidence of elevator-related accidents has not decreased. A study analyzing 48 elevator-related fatalities over a 30-year period across nine jurisdictions in the United States revealed that, while injuries and fatalities caused by elevators are rare, they encompass a diverse range of incidents, including suffocation, blunt force trauma, lacerations, and environmental injuries [6]. Those at risk of elevator-related fatalities include maintenance workers, construction workers, other staff, and users. Continued research into elevator installation techniques is vital to ensure the safety of these individuals.

### B. Study on Line Detection in Elevator Shafts

An examination of market trends in elevator installation reveals that in 2018, a Swiss elevator manufacturing company, known as Company S, introduced the first automated elevator shaft installation equipment known as R.I.S.E (Rapid Installation and Safe Erection)[7]. This equipment is designed to operate on only two of the four sides of the elevator shaft, where the robot deploys feet for stabilization before proceeding with the installation. It does not perform the process of detecting an absolute reference line to determine the robot's position[8]. In this paper, we conduct research on the detection of reference lines to determine the robot's absolute position, with the goal of developing equipment capable of operating on all four sides of the elevator shaft. Company S is the only entity to have developed automated elevator shaft installation equipment, and in response, our research team has partnered with Company H, the leading elevator manufacturer in South Korea, to initiate the first research and development of such automation equipment in Korea.

Previous studies related to this study primarily use the geometric features of lanes for detection and employ curve fitting methods. Recently, several automotive companies have introduced autonomous vehicles that utilize a variety of sensors, including vision sensors, radar, and LiDAR, to collect data. The collected data is then used to assess the driving environment and guide movement decisions[9]. This research involves lane detection using cameras, a type of vision sensor. The process includes steps such as video frame capture, bird's-eye view transformation, lane area extraction using HSV color values, noise removal with Gaussian filtering, edge extraction via Canny edge detection, and lane detection with the Hough transform. However, in the case of lane detection, the color of the lane typically differs significantly from that of the background, making it easier to distinguish. In contrast, within the elevator shaft, the color of the piano wire to be detected is nearly identical to the background color of the shaft walls, posing a significant challenge. Additionally, lanes are relatively thick, making it easier to extract color values from image data in pixels, whereas the piano wire in the elevator shaft is only about one pixel wide. In terms of brightness, lanes are well-illuminated, either by sunlight or vehicle headlights, whereas the piano wire in the elevator shaft is dimly lit by installed lighting, the extracted color values vary between the upper and lower parts of the target line. Consequently, detecting the line using existing research methods presents significant challenges.

This paper focuses on the detection of a reference wire to determine the position of a robot, with the goal of automating internal work within the elevator shaft during the elevator installation process. To enable the smooth operation of the auto installation robot within the shaft, the reference wire is set to a minimal thickness of 0.2 mm. The robot performs tasks on each floor of the elevator tower, detecting the reference wire and calculating the distance between the robot and the designated work location, where it carries out tasks such as anchor hole drilling or anchor bolt installation. The elevator shaft serves as the passageway for the elevator, and if work is performed outside the designated position, the elevator may not function properly. Therefore, accurately determining the position of the auto installation robot is crucial, and this paper seeks to achieve this by detecting the 0.2 mm piano wire as the reference line.

## III. METHODOLOGY

This study follows the process outlined in Figure 1. Section 'A. Image Calibration' covers the calibration of image data, which was directly collected for the purposes of this research. In section 'B. Image Denoising,' the study details the key preprocessing techniques applied, including Gaussian Blurring, Sharpening, Embossing, and Discrete Fourier Transform. Section 'C. Edge Detection' introduces the Canny Edge Detection method, while section 'D. Line Detection' describes the Hough Transform technique for line detection. Finally, in section 'E. Extracting Vertical Lines by Hough Averaging,' the study proposes a method to address the limitations of the Hough Transform when detecting straight lines.

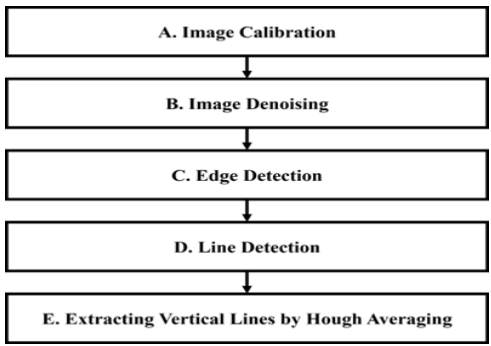

**FIGURE 1. Flowchart of the Methodology**

### A. IMAGE CALIBRATION

Image calibration refers to the process of determining parameter values during the projection of the three-dimensional real world onto a two-dimensional image plane[10]. The internal parameters, which are related to the optical characteristics of the camera lens, are encapsulated in the camera matrix, within includes details such as the camera focal length and the principal point. The camera focal length represents the distance between the lens center and the image sensor and is expressed in pixel units. The principal point is refers to the point where a perpendicular line dropped from the lens center intersects. In addition to internal parameters, distortion coefficients are also used during calibration. Lens distortion typically manifests in two forms: radial distortion and tangential distortion. Radial distortion causes pixels to become increasingly distorted as they move further from the center of the image. Specifically, barrel distortion, commonly seen in wide-angle lenses, causes the image to bulge outward from the center. Conversely, pincushion distortion, frequently occurring in telephoto lenses, results in the image pinching inward. Radial distortion is mathematically described by Equations 1 and 2 below, where r represents the distance from the center of the image, and k1, k2, and k3 are the radial distortion coefficients.

$$x_{distorted} = x \cdot (1 + k_1 \cdot r^2 + k_2 \cdot r^4 + k_3 \cdot r^6) \quad (1)$$

$$y_{distorted} = y \cdot (1 + k_1 \cdot r^2 + k_2 \cdot r^4 + k_3 \cdot r^6) \quad (2)$$

Tangential distortion occurs when the lens is not perfectly parallel, resulting in the image tilting in a specific direction. This type of distortion is mathematically modeled by Equations 3 and 4 below.

$$x_{distorted} = x + [2 \cdot p_1 \cdot xy + p_2 \cdot (r^2 + 2 \cdot x^2)] \quad (3)$$

$$y_{distorted} = y + [p_1 \cdot (r^2 + 2 \cdot y^2) + 2 \cdot p_2 \cdot xy] \quad (4)$$

The distortion coefficient array used in OpenCV typically follows the format shown in Equation 5. Here, k1, k2, and k3 are the radial distortion coefficients, while p1 and p2 represent the tangential distortion coefficients. k4, k5, and k6 are higher-order radial distortion coefficients, which are typically set to zero.

$$dist_{coeffs} = [k1, k2, p1, p2, k3, k4, k5, k6] \quad (5)$$

In the working environment of an elevator shaft, accurately detecting the reference wire is crucial for determining the position of the installation robot. By addressing the camera's distortion through image calibration, the actual position of the wire can be accurately estimated. In this paper, we utilize Python-based libraries to perform image calibration, leveraging information such as the camera's internal parameters and distortion coefficients. This method enhances the automated installation equipment's ability to function reliably with accurate positional data.

## B. IMAGE DENOISING

[Gaussian Blurring]

Gaussian blurring, a filtering technique used in image processing, is employed to reduce noise or blur details in an image[11]. It functions by smoothing out the pixel values within an image, removing extreme outliers that deviate from a uniform range. This effect is achieved using the Gaussian function, which is based on the normal distribution probability density function. The Gaussian function is described by Equation 6 below, where (x, y) represents the pixel coordinates, and σ is the standard deviation of the Gaussian function.

$$G(x,y) = \frac{1}{2\pi\sigma^2} e^{-\frac{x^2+y^2}{2\sigma^2}} \quad (6)$$

Gaussian blurring is applied to each pixel (x, y) in a given image by recalculating its value based on the Gaussian function applied to neighboring pixels. Weights are assigned according to the distance between pixels, derived from the Gaussian function. The calculated new value then replaces the original pixel value in the image. Greater weights are assigned to closer neighboring pixels, and a larger standard deviation σ results in a stronger blurring effect. However, excessive blurring can also obscure edges, reducing the accuracy of line detection.

Most edge detection algorithms are sensitive to noise, making it challenging to achieve accurate results[12]. The experimental environment in this study also contains significant noise, such as cracks in the walls of the elevator shaft or electrical wires, making it difficult to accurately detect the 0.2 mm reference wire. Therefore, it is necessary to remove background noise without damaging the target line. It has been observed that removing noise before edge detection improves the algorithm's performance[13]. Previous research has demonstrated the use of Gaussian blurring in studies focused on detecting edges or lines[14-17]. Gaussian blurring is effective in reducing noise by diminishing the high-frequency components of an image, without damaging the low-frequency components where edges reside, thereby improving the accuracy of edge detection algorithms. In this study, Gaussian blurring is performed using Python-based libraries.

[Sharpening]

Sharpening is a filtering technique used to enhance the clarity of an image by making the contours of objects more distinct. It is the opposite of blurring, as it accentuates the original pixels relative to the surrounding pixels, thereby emphasizing edge areas. The sharpening technique is typically implemented using a blurred version of the image, and in this process, an unsharp mask filter is employed. The operation of the unsharp mask filter is illustrated in Figure 2.

In Figure 2, the x-axis represents the movement of pixel coordinates, while the y-axis represents the pixel values. Figure 2(a) shows an increase in pixel values near the edge, and Figure 2(b) illustrates the result after applying blurring. Figure 2(c) represents the result obtained by subtracting the blurred version from the original, and in Figure 2(d), adding (a) and (c) produces the edge-enhanced image, as described by Equation 7. The sharpness can be adjusted by modifying the real-valued weight α.

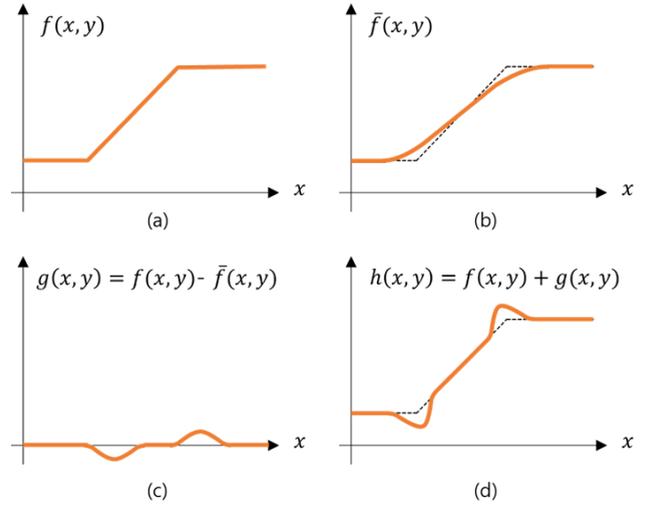

**FIGURE 2.** The Operation Mechanism of an Unsharp Mask Filter

$$h(x,y) = f(x,y) + \alpha \cdot g(x,y) \quad (7)$$

Sharpening is an important preprocessing method used to enhance image quality, particularly after noise removal, to improve feature point detection performance[18]. When only simple blurring is applied, the image can become overly smooth, causing the contours to blur and leading to difficulties in feature point detection. However, by additionally applying sharpening, these issues can be mitigated[19-21]. Sharpening is effective in clearly delineating edges within the image, making it particularly useful for distinguishing boundaries. It can also be employed to emphasize the vertical reference wire within the elevator shaft, allowing for more accurate feature point detection. However, since sharpening can also amplify noise, the noise removal stage is crucial. Excessive sharpening may result in incorrect edge detection, which could reduce the accuracy of detecting the reference wire. Therefore, careful adjustment of the sharpening process is necessary.

[Embossing]

Embossing is a technique used to impart a three-dimensional effect to an image. To apply the embossing effect, an n×n kernel (filter) is defined. The method works by calculating the differences between neighboring pixels relative to the center of the kernel. After converting the image to grayscale, the kernel is applied to each pixel of the image to adjust the brightness values, thereby creating a three-dimensional effect. The equation for applying a 5x5 embossing filter $K$ to the input image $I(x,y)$ is given by Equation 8 below. Here, $K(i,j)$ represents each element of the embossing filter, and , $I(x+i, y+j)$ denotes the surrounding pixel values of the input image. The boundary areas of objects are set to be brighter or darker based on relatively flat regions with minimal pixel value changes. In areas where pixel values change sharply in the diagonal direction, the pixel values become much greater or less than zero, enhancing the three-dimensional appearance.

$$E(x,y) = \sum_{i=-1}^{1}\sum_{j=-1}^{1} K(i,j) \cdot I(x+i, y+j) \qquad (8)$$

The embossing technique is utilized to more distinctly emphasize object boundaries and add depth to images, making it useful in studies focused on detecting objects or boundaries within images[22-23]. In this study, the embossing technique is also applied to enhance the detection rate of the reference wire inside the elevator shaft. However, it is important to note that embossing can create artificial edges, potentially emphasizing noise in addition to the reference wire. Therefore, caution must be exercised when using this technique.

[Discrete Fourier Transform]

Another technique for denoising images is the Fourier transform. The Fourier transform, proposed by Joseph Fourier, is an algorithm that decomposes an arbitrary input signal into a sum of periodic functions, each with different frequencies. When applied to image processing, the Fourier transform converts the spatial information of an image into the frequency domain, allowing for the analysis of the frequency components within the image. Spatial frequency refers to the number of times a particular pattern of pixel variation repeats per unit of distance in the image. By filtering these frequencies, the image can be made either blurrier or sharper. Low frequencies correspond to areas with little pixel variation, while high frequencies correspond to areas with significant pixel variation. Filtering out low frequencies allows only the low-frequency components to pass through, resulting in blurring the edges. Conversely, filtering out high frequencies sharpens the edges by allowing only the high-frequency components to pass through. In this study, high-frequency filtering is applied to the Fourier transform to enhance edges, as edge detection is the primary objective.

Since images are two-dimensional, both the x-axis and y-axis variations must be considered, and the signals are discrete rather than continuous. Therefore, the discrete Fourier transform for an image of size W x H is represented by Equation 9 below. Here, $j$ is the imaginary unit ($j = \sqrt{-1}$), f(x, y) represents the original input signal, and $e^{j2\pi(ux/W+vy/H)}$ represents the periodic function component with x-axis frequency u/W and y-axis frequency v/H that constitutes f(x, y). In Equation 10, F(u, v) is the coefficient of the periodic function component with frequencies u/W and v/H, representing the amplitude of the periodic function components.

$$f(x,y) = \sum_{u=0}^{W-1}\sum_{v=0}^{H-1} F(u,v) e^{j2\pi(ux/W+vy/H)} \qquad (9)$$

$$F(u,v) = \frac{1}{WH}\sum_{x=0}^{W-1}\sum_{y=0}^{H-1} f(x,y) e^{-j2\pi(ux/W+vy/H)} \qquad (10)$$

The discrete Fourier transform is a technique used to analyze the frequency characteristics of digital signals, such as images or audio signals, and is also employed for tasks like filtering, compression, and synthesis. In image processing, particularly, high-frequency filtering in the frequency domain enhances the edges, which are areas with significant pixel value variations, thereby improving edge detection and noise reduction performance[24-26]. This study leverages these characteristics to enhance the detection rate of the reference wire within the elevator shaft. While the Fourier transform is highly effective at enhancing edges and removing noise, it is computationally intensive and may result in the loss of details in the original image. However, since the primary goal in this study is to accurately detect the reference line, the potential loss of detail in the original image is of minimal concern.

### C. EDGE DETECTION
[Canny Edge Detection]

Canny Edge Detection, developed by John F. Canny in 1986, is a widely used detection algorithm in the fields of computer vision and image processing[27]. The Canny Edge Detection technique is robust to noise and can accurately detect edge positions through non-maximum suppression. Additionally, it employs a dual-threshold approach to classify potential edges into strong edges and weak edges, making it a reliable method. Due to these advantages, the Canny edge detection technique is extensively utilized in edge detection research[28-30].

To perform Canny edge detection, noise is first removed from the image. Then, the Sobel operator is used to calculate the gradient of the image along the x and y axes, determining the gradient magnitude and direction for each pixel. Non-maximum suppression is applied to retain only the pixels with the highest gradient magnitude, thereby forming thin edges. A dual-threshold is then applied to classify the formed edges into strong edges, which exceed the high threshold, and weak edges, which fall between the low and high thresholds. Only weak edges connected to strong edges are considered as final edges.

While the Canny Edge Detection technique can be overwhelmed by a large number of detected edges in images with complex textures, and it is sensitive to changes in lighting conditions, that may yield varying results for the same object under different lighting. However, the images used in this study are relatively simple, with minimal texture complexity, and the experimental conditions allow for consistent lighting, enabling effective use of the Canny detection technique.

### D. LINE DETECTION
[Hough Transform]

The Hough transform, proposed by F.V.C. Hough in 1962, is a technique used in image processing and computer vision for detecting specific shapes within an image[31]. In this study, we aim to detect straight lines using the edge information obtained from the Canny Edge Detection in Section 3.3, applying the Hough transform technique. The Hough transform detects lines by converting the line equation from the 2D xy-coordinate space to the parameter space. While a line is typically represented by the equation $y = mx + b$ with slope $m$ and intercept $b$, the slope $m$ can approach infinity, which poses challenges. To address this, the Hough transform converts the equation into polar coordinates, representing the line as $\rho = x\cos\theta + y\sin\theta$, where $\rho$ is the distance from the origin to the line and $\theta$ is the angle between the line and the x-axis. For each point in the image, all possible lines passing through that point are represented in the parameter space by varying $\theta$ and calculating $\rho = x\cos\theta + y\sin\theta$. The results are accumulated in the parameter space, and the line is detected by identifying the maximum value in the accumulator array.

The Hough transform is well-suited for detecting very thin lines while minimizing the effects of noise. Because it detects lines by finding the maximum value in the accumulator array, it is robust even when some data is missing or incomplete. Consequently, the Hough transform is frequently used in studies focused on line and edge detection[32-35]. In this paper, we apply the Hough transform to minimize noise in the elevator shaft installation environment and accurately detect the reference wire.

### E. EXTRACTING VERTICAL LINES BY HOUGH AVERAGING

This study aims to determine the position of an installation robot inside the elevator shaft by using a camera to detect a thin 0.2 mm piano wire as a reference line. Due to the extremely thin nature of the wire, the Hough transform alone, even after denoising, can result in the surrounding noise being mistakenly detected along with the desired line. Additionally, the reference line may not be detected as a single continuous line, but rather as multiple short, fragmented lines.

To address these issues, our research team introduces several modifications to the Hough transform. We first assume that the piano wire being detected is vertical. Consequently, the detection process is conditioned to identify only vertical lines. Furthermore, by extracting the x-coordinates of the detected lines, we can merge those with identical x-coordinates, thereby resolving the issue of fragmented line detection and allowing the reference line to be detected as one continuous line. The detailed process is as follows:

First, the x-coordinates of the lines detected by the Hough transform are stored in a list, and the range in which the x-coordinates are most densely clustered is identified for each image. In this study, we calculate the mean $\bar{x}$ of the x-coordinates within the range where the most x-coordinates are located, within a 20-pixel range. A vertical line is then drawn at the averaged $\bar{x}$ value to represent the reference line.

## IV. EXPERIMENT

In this study, we aim to conduct a comparative experiment on four different algorithms that follow distinct preprocessing procedures to detect a thin 0.2 mm piano wire. Experiment 1 (GCH) involves applying Gaussian blurring as the preprocessing step, followed by Canny edge detection and the Hough transform to detect straight lines. Experiment 2 (GSCH) applies both Gaussian blurring and sharpening techniques in the preprocessing stage, followed by Canny edge detection and the Hough transform. Experiment 3 (GECH) involves the application of Gaussian blurring and embossing techniques as preprocessing, followed by Canny edge detection and the Hough transform. Experiment 4 (FCH) applies the Fourier transform as the preprocessing step, followed by Canny edge detection and the Hough transform to detect straight lines. Chapter 4 compares these four experiments, as summarized in Table I and Figure 3, to propose the most effective method for detecting the 0.2 mm piano wire.

TABLE I
EXPERIMENTAL STUDY

|  | Preprocessing | Edge Detection | Line Detection |
|---|---|---|---|
| Experiment 1 (GCH) | **G**aussian **B**lurring | **C**anny Edge Detection | **H**ough Transform |
| Experiment 2 (GSCH) | **G**aussian **B**lurring + **S**harpening | | |
| Experiment 3 (GECH) | **G**aussian **B**lurring + **E**mbossing | | |
| Experiment 4 (FCH) | **F**ourier Transform | | |

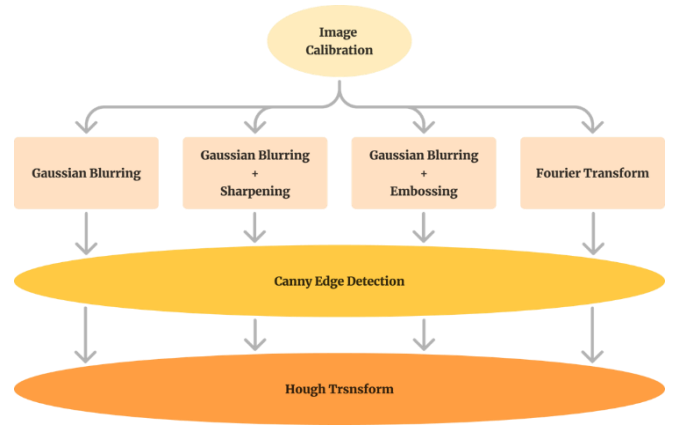

**FIGURE 3.** Experimental Environment

## A. DATA COLLECTING FOR LINE DETECTION

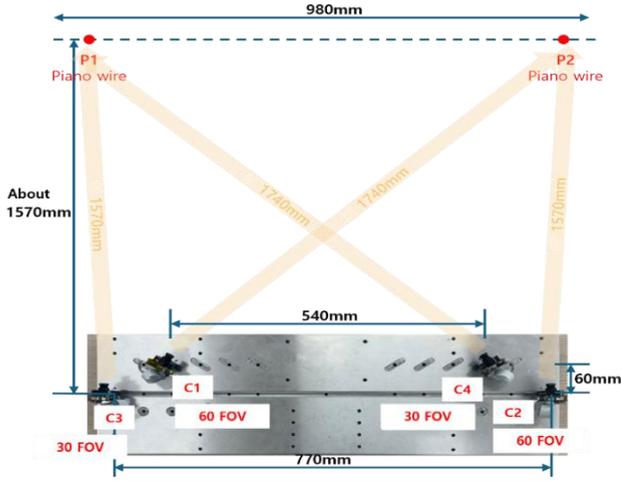

**FIGURE 4.** Experimental Environment

The environment used for data collection in this study is illustrated in Figure 4. This experiment aims to calculate the position of a robotic system operating inside an elevator shaft using camera images captured from a long-distance diagonal perspective (C4, C1) and a short-distance straight perspective (C3, C2) for two reference lines (P1, P2). Four lenses (C1, C2, C3, C4) are employed to detect two piano wires (P1, P2), with the lenses having two different fields of view: 30 degrees and 60 degrees. As shown in Figure 4, the C1 and C2 lenses capture P2 from diagonal and straight directions, respectively, while the C3 and C4 lenses capture P1 from straight and diagonal directions, respectively. The distance between the lenses and the reference line is approximately 1570 mm for the straight direction and 1740 mm for the diagonal direction, which aligns with the specifications of the experimental tower from Company H.

The four lenses were mounted on a plywood board that was attached to a makeshift elevator. The data collection process began when the elevator was stationary at the 5th floor, and images were captured as the elevator descended to the 1st floor and then ascended back to the 5th floor, with the shaft images being recorded at 5 frames per second. Images captured by C3 of P1 are labeled as LtoL, those by C1 of P2 as LtoR, those by C4 of P1 as RtoL, and those by C2 of P2 as RtoR. The four datasets were collected by capturing reference lines P1 and P2 from different angles (30° and 60° field of view) and distances (1570 mm and 1740 mm), thereby diversifying the environmental conditions for reference line detection. The number of images collected was 828 in the first round, 801 in the second round, 805 in the third round, and 811 in the fourth round. The slight variations in the number of images collected in each round are due to the exclusion of duplicate images captured when the elevator was stationary, as duplicate images captured during stationary periods were not included in the research data.

## B. IMAGE CALIBRATION

For the purpose of image calibration, this study conducted the calibration process with the assistance of a camera expert collaborating with our research team. The internal camera parameters (Camera Focal Length, Principal Point) and distortion coefficients (Radial Distortion Coefficients, Tangential Distortion Coefficients) used in the calibration are presented in Table II. The higher-order radial distortion coefficients k4, k5, and k6 were all set to zero.

TABLE II
CAMERA PARAMETERS AND DISTORTION COEFFICIENTS

|  |  | Camera 1(C1) | Camera 2(C2) | Camera 3(C3) | Camera 4(C4) |
|---|---|---|---|---|---|
| Camera Focal Length | fx | 3859.63 | 3786.44 | 3812.58 | 3811.30 |
|  | fy | 3853.00 | 3793.20 | 3824.67 | 3829.32 |
| Principle Point | cx | 1988.85 | 1939.22 | 1887.29 | 1945.61 |
|  | cy | 1469.96 | 1564.98 | 1544.79 | 1647.71 |
| Radial Distortion Coefficients | k1 | -0.4883 | -0.5402 | -0.5551 | -0.4865 |
|  | k2 | -0.5526 | 0.0792 | 0.2221 | -0.4850 |
|  | k3 | 3.6464 | 0.8875 | 0.2079 | 2.9886 |
| Tangential Distortion Coefficients | p1 | 0.0042 | 0.0031 | 0.0036 | -0.0008 |
|  | p2 | 0.0001 | 0.0031 | 0.0059 | 0.0025 |

Based on the information in Table II, the distorted images are corrected using the 'cv2.undistort()' function. Subsequently, the Region of Interest (ROI) is applied to crop the valid area of the corrected image, and the resulting image is saved. Figure 5 provides an example from the image dataset used in this study, showing a comparison of the images before and after calibration.

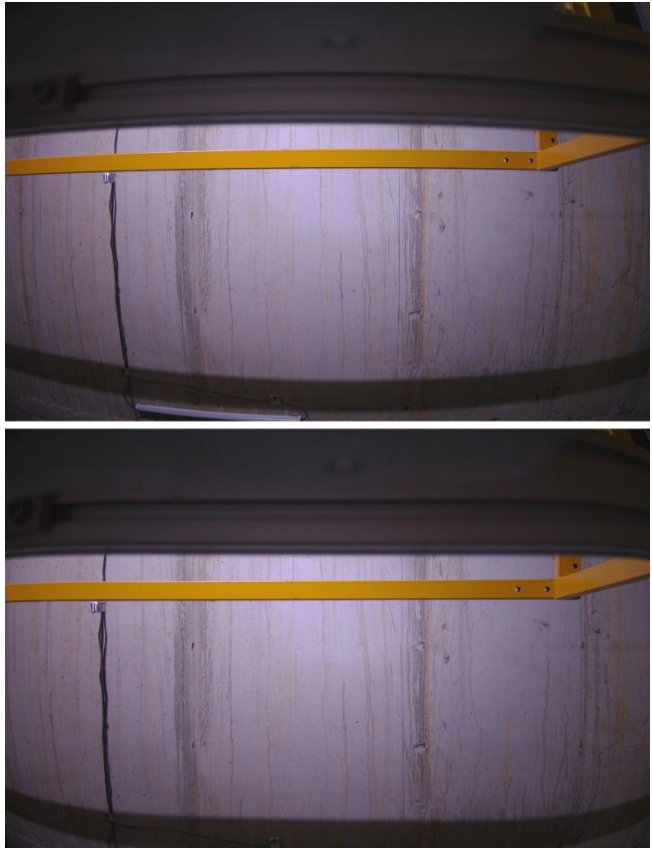

**FIGURE 5.** The Result of Calibration (Top: Before, Bottom: After)
Bottom: After)

### C. IMAGE DENOISING

The thresholds for the four experiments conducted in this study are presented in Tables III, IV, V, and VI. All experiments involve preprocessing followed by Canny edge detection and the Hough transform to detect the 0.2 mm piano wire. In Experiment 1 (GCH), only Gaussian blurring was performed during preprocessing. Experiment 2 (GSCH) involved both Gaussian blurring and sharpening, while Experiment 3 (GECH) applied Gaussian blurring and embossing. Experiment 4 (FCH), which produced the most outstanding results in this study, utilized the Fourier transform. The optimal thresholds, yielding the highest detection rates, were determined through multiple iterations of each experiment and applied accordingly. For instance, in the LtoL data shown in Table III, a (5, 5) Gaussian blur filter achieved the highest detection rate in both Experiment 1 (GCH) and Experiment 3 (GECH), while a (7, 7) filter was optimal for Experiment 2 (GSCH), and these filters were used in the line detection experiments. Tables III, IV, V, and VI correspond to the experimental thresholds for LtoL, LtoR, RtoL, and RtoR, respectively.

TABLE III
EXPERIMENTAL THRESHOLDS FOR THE PREPROCESSING OF LtoL DATA

| | | Experiment | | | |
|---|---|---|---|---|---|
| | | Exp1 (GCH) | Exp2 (GSCH) | Exp3 (GECH) | Exp4 (FCH) |
| Preprocessing | Gaussian Blurred Filter | (5, 5) | (7, 7) | (5, 5) | - |
| | Sharpening Center Pixel | - | 5 | - | - |
| | Emboss Filter | - | - | -5 -4 0 / -4 -2 4 / 0 4 5 | - |
| | Fourier Mask Radius | - | - | - | 500 |

TABLE IV
EXPERIMENTAL THRESHOLDS FOR THE PREPROCESSING OF LtoR DATA

| | | Experiment | | | |
|---|---|---|---|---|---|
| | | Exp1 (GCH) | Exp2 (GSCH) | Exp3 (GECH) | Exp4 (FCH) |
| Preprocessing | Gaussian Blurred Filter | (5, 5) | (7, 7) | (5, 5) | - |
| | Sharpening Center Pixel | - | 5 | - | - |
| | Emboss Filter | - | - | -5 -4 0 / -4 -2 4 / 0 4 5 | - |
| | Fourier Mask Radius | - | - | - | 40 |

TABLE V
EXPERIMENTAL THRESHOLDS FOR THE PREPROCESSING OF RtoL DATA

| | | Experiment | | | |
|---|---|---|---|---|---|
| | | Exp1 (GCH) | Exp2 (GSCH) | Exp3 (GECH) | Exp4 (FCH) |
| Preprocessing | Gaussian Blurred Filter | (5, 5) | (7, 7) | (5, 5) | - |
| | Sharpening Center Pixel | - | 5 | - | - |
| | Emboss Filter | - | - | -5 -4 0 / -4 -2 4 / 0 4 5 | - |
| | Fourier Mask Radius | - | - | - | 400 |

TABLE VI
EXPERIMENTAL THRESHOLDS FOR THE PREPROCESSING OF RTOR DATA

|  |  | Experiment | | | |
| --- | --- | --- | --- | --- | --- |
|  |  | Exp1 (GCH) | Exp2 (GSCH) | Exp3 (GECH) | Exp4 (FCH) |
| Preprocessing | Gaussian Blurred Filter | (5, 5) | (7, 7) | (5, 5) | - |
|  | Sharpening Center Pixel | - | 5 | - | - |
|  | Emboss Filter | - | - | -5 -4 0 <br> -4 -2 4 <br> 0 4 5 | - |
|  | Fourier Mask Radius | - | - | - | 200 |

### D. EDGE DETECTION

Based on the thresholds in Tables III, IV, V, and VI, edges are detected using the Canny detection method. The thresholds for Canny Edge Detection applied to the LtoL, LtoR, RtoL, and RtoR datasets are presented in Table VII.

TABLE VII
EXPERIMENTAL THRESHOLDS FOR CANNY EDGE DETECTION

|  |  | Experiment | | | |
| --- | --- | --- | --- | --- | --- |
|  |  | Exp1 (GCH) | Exp2 (GSCH) | Exp3 (GECH) | Exp4 (FCH) |
| Canny | LtoL Data | [1, 100] | [1, 100] | [1, 100] | [1, 100] |
|  | LtoR Data | [1, 100] | [1, 100] | [1, 100] | [1, 100] |
|  | RtoL Data | [1, 100] | [1, 100] | [1, 100] | [1, 100] |
|  | RtoR Data | [1, 100] | [1, 100] | [1, 100] | [1, 100] |

Figure 6 provides an example from the image dataset used in this study, showing a comparison of the images before and after performing Canny edge detection.

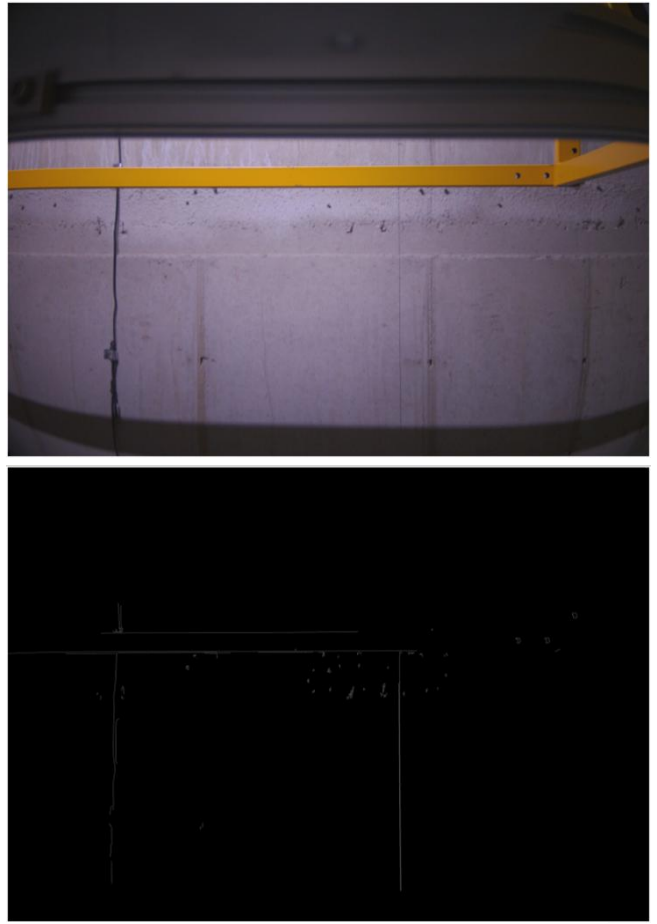

**FIGURE 6. The Result of Canny Edge Detection (Top: Before, Bottom: After)**

Non-edge areas are converted to black, while the detected edges are displayed in white. The dataset used in this study includes noise elements such as cracks in the inner walls of the elevator shaft, additional electric wires, and safety bars used during filming. However, the edge-detected images reveal only the distinct vertical and horizontal lines from the original image, demonstrating that even the extremely thin 0.2 mm wire is detected without difficulty. The edge detection results provide a robust foundation for subsequent line detection using the Hough transform.

### E. LINE DETECTION

Based on the edges detected in Section IV.D, the Hough transform is employed to detect the 0.2 mm piano wire. The experimental thresholds for Line Detection were determined by conducting multiple iterations for each experiment (GCH, GSCH, GECH, FCH) to identify the optimal values. For the LtoL, LtoR, RtoL, and RtoR datasets, the Hough Thresholds were set to 100, and the Maximum Line Gap was set to 10 for all experiments. The optimal threshold for Minimum Line Length was set to 50 for LtoL and 100 for the remaining datasets. The Experimental Thresholds for Line Detection are presented in Table VIII.

TABLE VIII
EXPERIMENTAL THRESHOLDS FOR LINE DETECTION

|  |  | Experiment 1, 2, 3, 4 (GCH, GSCH, GECH, FCH) | | | |
|---|---|---|---|---|---|
| Minimum Line Length | LtoL Data | 50 | 50 | 50 | 50 |
|  | LtoR Data | 160 | 160 | 160 | 160 |
|  | RtoL Data | 50 | 50 | 50 | 50 |
|  | RtoR Data | 150 | 150 | 150 | 150 |
| Maximum Line Gap | All Data | 10 | 10 | 10 | 10 |
| Hough Thresholds | All Data | 100 | 100 | 100 | 100 |

Figure 7 provides an example from the image dataset used in this study, showing a comparison of the images before and after performing the Hough Transform for piano wire detection.

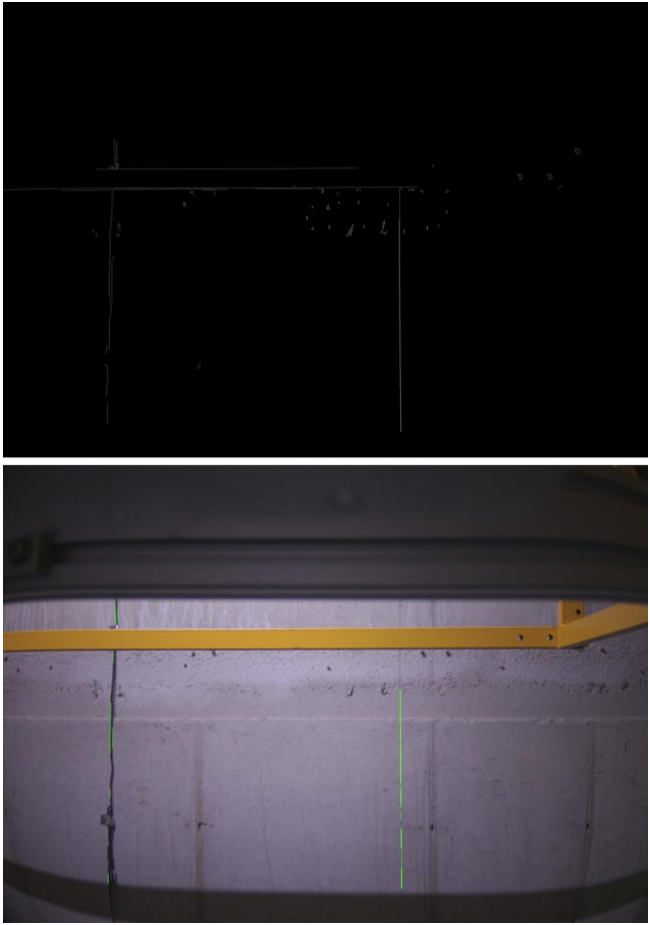

**FIGURE 7.** The Result of Hough Line Detection (Top: Before, Bottom: After)

## F. EXTRACTING VERTICAL LINES BY HOUGH AVERAGING

After performing various preprocessing steps, such as Gaussian blurring and Fourier transform, followed by Canny edge detection and the Hough transform, additional computations were carried out to accurately extract only the piano wire (reference line). The x-coordinates of the lines detected by the Hough transform were stored in a list. For each image, the x-coordinates within a 20-pixel range where the x-coordinates were most densely clustered were averaged to calculate the $\bar{x}$ value. A vertical line was then drawn at the $\bar{x}$ value to represent the reference line.

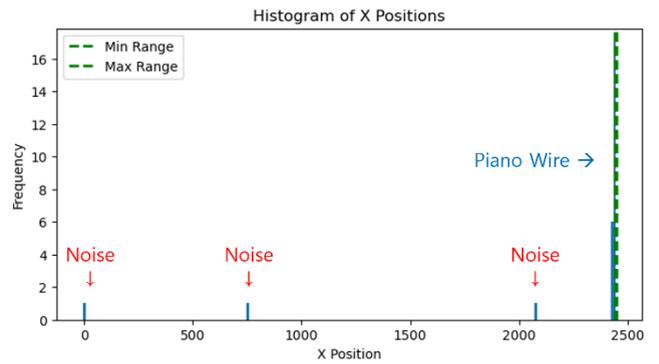

**FIGURE 8.** The Example of Extracting Vertical Lines by Hough Averaging

As shown in Figure 8, lines that appear to be noise are detected near the x-coordinates of 0, 700, and 2100. This computation not only eliminates such noise but also addresses the issue of detecting a single piano wire as multiple short lines. Through this process, the 0.2 mm piano wire intended for detection in this study is accurately identified. An example of the resulting detected image is presented in Figure 9 below.

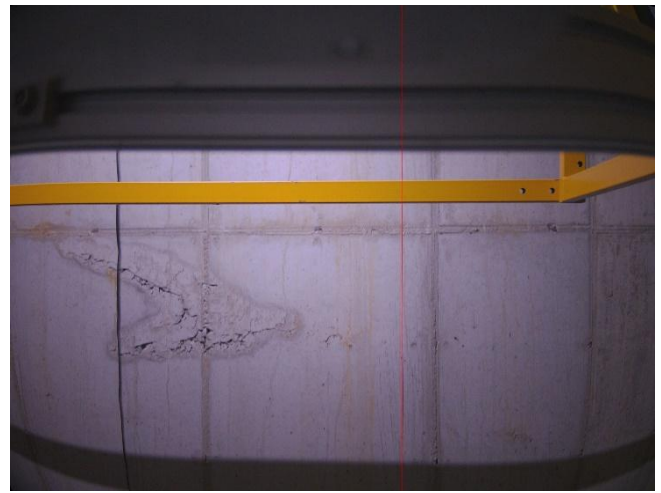

**FIGURE 9.** LtoR Data Image

## V. Experimental Results and Discussion

The experimental results for the LtoL, LtoR, RtoL, and RtoR datasets, based on the experiments in Chapter IV, are summarized in Table IX.

TABLE IX
COMPARISON OF EXPERIMENTAL ACCURACY

|  |  | Accuracy | | | |
|---|---|---|---|---|---|
|  |  | Exp1 (GCH) | Exp2 (GSCH) | Exp3 (GECH) | Exp4 (FCH) |
| LtoL | First Session | 84.04% | 40.02% | 34.22% | **99.52%** |
|  | Second Session | 85.75% | 59.62% | 5.88% | **99.75%** |
|  | Third Session | 85.20% | 78.98% | 1.00% | 0.50% |
|  | Fourth Session | 73.09% | 31.85% | 18.02% | **99.38%** |
| LtoR | First Session | 36.28% | 48.37% | 0.00% | **59.37%** |
|  | Second Session | 32.00% | 44.38% | 0.00% | **49.50%** |
|  | Third Session | 8.33% | 4.98% | 0.00% | 0.12% |
|  | Fourth Session | 39.75% | 52.47% | 0.00% | **61.98%** |
| RtoL | First Session | 81.14% | 60.58% | 1.21% | **87.79%** |
|  | Second Session | 90.00% | 78.75% | 2.38% | **92.38%** |
|  | Third Session | 46.77% | 46.02% | 0.87% | 21.64% |
|  | Fourth Session | 78.15% | 58.15% | 0.99% | **89.14%** |
| RtoR | First Session | 98.31% | **99.15%** | 6.53% | 84.64% |
|  | Second Session | 96.13% | **98.00%** | 6.38% | 71.38% |
|  | Third Session | 86.70% | 52.11% | 0.00% | 16.67% |
|  | Fourth Session | 98.89% | **99.14%** | 4.57% | 89.26% |

The experimental results indicate that Experiment 4, which employed the Fourier transform as the preprocessing method, achieved the highest line detection rate for the LtoL, LtoR, and RtoL datasets. In contrast, Experiment 2, which utilized Gaussian blurring and sharpening, yielded the highest line detection rate for the RtoR dataset. Experiment 1 uses a basic preprocessing method, but in environments with high noise, edge information may be insufficient, leading to decreased baseline detection performance. Experiment 2 applies a sharpening technique that emphasizes edges, making the baseline more distinct. In the case of the RtoR dataset, the baseline was detected well across all three experiments (1, 2, and 4), with Experiment 2 showing particularly high performance. This can be attributed to the enhanced edge emphasis, which made the baseline more prominent. Experiment 3 employs embossing, which enhances contrast, but this also highlights texture information, causing unnecessary patterns to emerge and lowering detection performance. Experiment 4, which uses the Fourier transform, removes noise in the frequency domain and emphasizes specific straight lines. This method is particularly effective when there is significant surrounding noise, such as wall cracks and wires, as it removes random noise while highlighting the periodic baseline, leading to the highest performance. However, for the RtoR dataset, which lacks external noise like wall cracks or wires, the benefits of the Fourier technique were not fully realized, and the performance of Experiment 2, which applied sharpening to more strongly emphasize the existing edges, was superior. Despite this, with the current line detection rate ranging from 70% to 89%, further data collection and algorithm refinement could lead to performance improvements. Additionally, it was noted that the third session of Experiment 4 (FCH), which used the Fourier transform as the preprocessing technique, had a significantly lower detection rate. This was likely due to improper lighting conditions during filming, which resulted in an environment that was too dark for accurate line detection. As shown in Figure 10, when comparing the third session with the others, it is evident that light reflection off the camera mounting equipment caused a noticeable difference, particularly in Experiment 4, where brightness-based preprocessing was applied, leading to the lower detection rate.

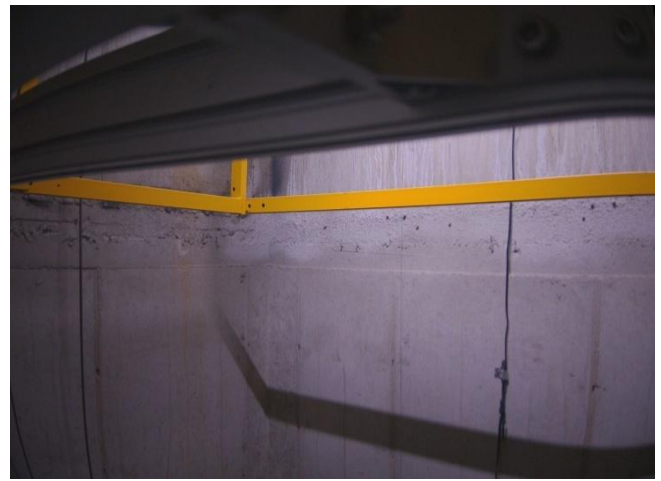

**FIGURE 10.** LtoR Data Image

In the case of Experiment 3 (GECH), the detection rate was lower across all datasets compared to the other experiments. This appears to be due to the nature of the embossing technique, which blurs the image during preprocessing, causing the target line to blur as well. Additionally, unlike the other datasets that generally achieved a line detection rate of over 90%, the LtoR dataset showed a significantly lower detection rate of around 50-60%. This reduced detection rate for LtoR is likely due to the inclusion of the elevator's moving ropes in the image (Figure 11).

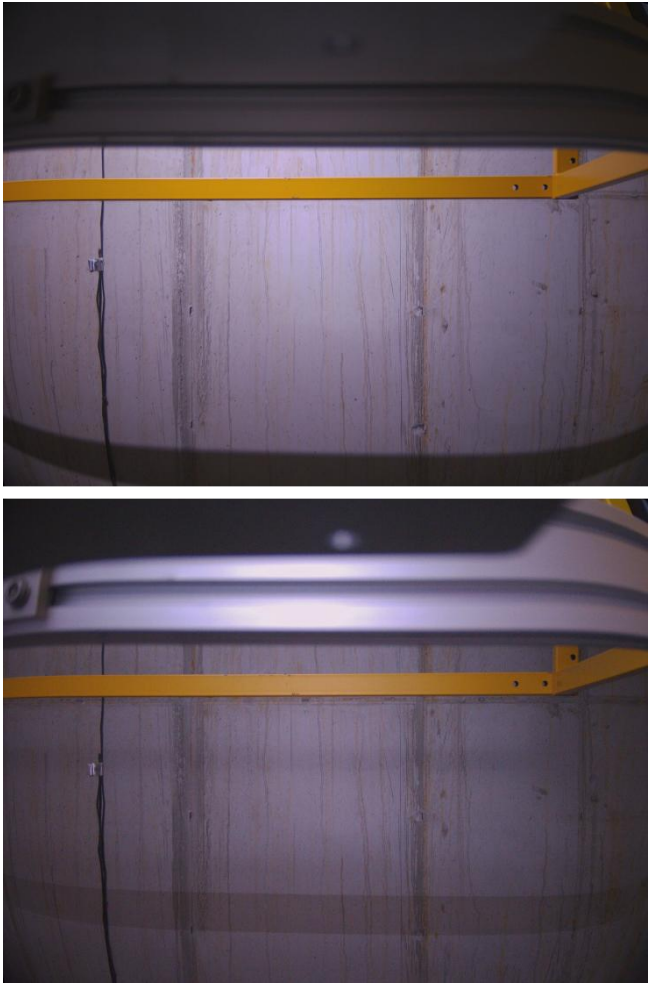

**FIGURE 11.** Differences in Shooting Episode's Images (Top: First Session, Bottom: Third Session)

The study results confirm that the 0.2 mm piano wire, serving as the reference line, can be detected at a high level despite the presence of significant noise, such as additional electric wires, wall cracks, and elevator ropes. The experiment that utilized the Fourier transform as a preprocessing step, Experiment 4, demonstrated superior detection rates, and even in the RtoR case, a generally high detection rate was achieved.

## VI. Conclusion

This study focuses on the detection of reference lines for determining the position of an automated installation robot within an elevator shaft. We developed an algorithm capable of detecting a thin 0.2 mm piano wire, which does not interfere with the working range of the automated installation robot. To develop this wire detection algorithm, our research team collected data directly from the experimental tower of H Corporation, the leading company in the industry, by capturing images of two wires from four different angles over four sessions. This study aims to calculate the position of a robotic system operating inside an elevator shaft using camera images captured from a long-distance diagonal perspective (C4, C1) and a short-distance straight perspective (C3, C2) for two reference lines (P1, P2). By enhancing the reference line detection performance, this study seeks to contribute to the automation of robotic operations within elevator shafts. A total of 3,245 images were collected, with 828 images in the first session, 801 in the second, 805 in the third, and 811 in the fourth for each camera angle. With four camera angles used, the entire image dataset constructed for line detection totaled 12,980 images.

We conducted comparative experiments using four different approaches (GCH, GSCH, GECH, FCH) and addressed the issue in the Hough transform of detecting a single line as multiple short lines, thereby improving line detection accuracy. As a result of this study, a high line detection rate of over 90% was achieved. The computation time for each experiment, on average per image, is as follows: 0.0929 seconds for GCH, 0.1002 seconds for GSCH, 0.0979 seconds for GECH, and 1.8541 seconds for FCH. The computer specifications used in the experiment are as follows: Apple M1 chip, 8-core CPU, 7-core GPU, 16-core Neural Engine, 8GB of RAM, and Python version 3.12.4.

This study enables the accurate calculation of location and control of tasks for an automated robot installed inside an elevator shaft. It is highly versatile and can be applied to any work environment requiring reference line recognition. Moreover, as the study focuses on recognizing a very thin line of 0.2 mm, the proposed methodology is applicable to various environments in which robots perform tasks, including constrained workspaces such as tunnels or subway tracks. This study is also significant in that it develops a recognition algorithm based on the straight-line properties of the line itself, rather than relying on RGB or HSV color values for line recognition. By utilizing this methodology, it is possible to accurately recognize the line even when the color of the upper and lower parts of the target line differs due to lighting, and the line can still be detected even if its thickness is extremely thin, visible to the naked eye. Additionally, recent case studies on edge and line detection have emerged, and this study can serve as a valuable precursor for future research in the [36, 37].

Among the four experiments conducted, the FCH (Fourier Canny Hough) Algorithm achieved relatively higher accuracy. This is likely due to the characteristic of the Fourier Transform that allows for the removal of noise without damaging the low-frequency boundary areas, making it effective in detecting the very thin 0.2 mm line targeted in this study. Since the current data was collected in a simulated experimental environment for baseline data acquisition, the inclusion of the elevator's moving ropes in some images has led to a decrease in baseline recognition performance. In a real-world environment, where the elevator's moving ropes are absent, an improvement in line detection performance is expected. However, to enhance reliability, additional images should be collected while ensuring that the elevator's moving ropes are excluded from the frame. Additionally, certain sessions exhibited lower detection rates than others, likely due to noise and inconsistencies in lighting conditions. Adjusting the lighting conditions and the camera's field of view during image acquisition, along with incorporating these modifications into the algorithm, is expected to further improve performance. Addressing these issues in future research could significantly enhance the overall robustness of this study. Furthermore, the methods employed in this study can be further optimized by fine-tuning the threshold values. In this study, thresholds were determined using a random search method, which is more time-efficient compared to grid search. As future work, refining these threshold values could contribute to the development of a more advanced line recognition algorithm. Therefore, our research team intends to supplement the algorithm by collecting additional data and will conduct further studies aimed at developing a machine learning-based line detection algorithm that automatically optimizes hyperparameters.

## ACKNOWLEDGMENT

This work was supported by the Technology Innovation Program (20023907, Development of Automation Equipment for Elevator Hoistway Installation) funded By the Ministry of Trade, Industry & Energy(MOTIE, Korea)

## REFERENCES

[1] Papanikolaou, N., Loupis, M., Spiropoulos, N., Mitronikas, E., Tatakis, E., Christodoulou, C., ... & Tsiftsis, T., "On the investigation of energy saving aspects of commercial lifts," Energy Efficiency, vol. 10, pp. 945-956, 2017.
[2] Zarikas, V., Loupis, M., Papanikolaou, N., & Kyritsi, C., "Statistical survey of elevator accidents in Greece," Safety science, vol. 59, pp. 93-103, 2013.
[3] Siikonen, M. L., "Current and future trends in vertical transportation," European Journal of Operational Research, 2024.
[4] Gołuch, P., Kuchmister, J., Ćmielewski, K., & Bryś, H., "Multi-sensors measuring system for geodetic monitoring of elevator guide rails," Measurement, vol. 130, pp. 18-31, 2018.
[5] Vrochidis, A., Charalampous, P., Dimitriou, N., Kladovasilakis, N., Chatzakis, M., Georgiadis, G., Tzovaras, D., and Krinidis, S., "Automatic elevator shaft inspection using a multi-sensor measuring system and computer vision techniques," Journal of Building Engineering, vol. 82, pp. 108358, 2024.
[6] Prahlow, J. A., Ashraf, Z., Plaza, N., Rogers, C., Ferreira, P., Fowler, D. R., ... & Lantz, P. E., "Elevator-related deaths," Journal of forensic sciences, vol. 65 no. 3, pp. 823-832, 2020.
[7] Schindler, Council on Tall Buildings and Urban Habitat(CTBUH) Conference, Dubai & Abu Dhabi, UAE, 2018.
[8] Hofer Philip, Urs Püntener, and Miguel Castro, "Technology Redefining the Future of Elevator Installation Methods for High Rise Buildings," 14th Symposium on Lift & Escalator Technologies, Vol. 14, No. 1, pp. 91-101, 2023.
[9] Xing, Y., Lv, C., Chen, L., Wang, H., Wang, H., Cao, D., and Wang, F. Y., "Advances in vision-based lane detection: Algorithms, integration, assessment, and perspectives on ACP-based parallel vision," IEEE/CAA Journal of Automatica Sinica, vol. 5, no. 3, pp. 645-661, 2018.
[10] Lee, S., Jeong, S., Yu, H., Kim, G., Kwak, H., Kang, E., and Lee, S., "Efficient image transformation and camera registration for the multi-projector image calibration," TECHART: Journal of Arts and Imaging Science, vol. 3, no. 1, pp. 38-42, 2016.
[11] Devi, T. G., Patil, N., Rai, S., and Philipose, C. S., "Gaussian blurring technique for detecting and classifying acute lymphoblastic leukemia cancer cells from microscopic biopsy images," Life, vol. 13, no. 2 pp. 348, 2023.
[12] Pimpalkhute, V. A., Page, R., Kothari, A., Bhurchandi, K. M., and Kamble, V. M., "Digital image noise estimation using DWT coefficients," IEEE transactions on image processing, vol. 30, pp. 1962-1972, 2021.
[13] Tsai, M. H., Fan-Jiang, J. C., Liou, G. Y., Cheng, F. J., Hwang, S. J., Peng, H. S., and Chu, H. Y., "Development of an online quality control system for injection molding process," Polymers, vol. 14, no. 8, pp. 1607, 2022.
[14] Xie, D., Wu, J., Zhao, B., & Pang, W., An Improved Gaussian-Mapping Algorithm for Edge Detection Based on FPGA, 2023 9th International Conference on Systems and Informatics (ICSAI), IEEE, 2023.
[15] An, Y., Jing, J., & Zhang, W., "Edge detection using multi-directional anisotropic Gaussian directional derivative," Signal, Image and Video Processing, vol. 17, no.7, pp. 3767-3774, 2023.
[16] Patwari, B., Nandi, U., & Ghosal, S. K., "Image steganography based on difference of Gaussians edge detection," Multimedia Tools and Applications, vol. 82, no. 28, pp. 43759-43779, 2023.
[17] Wai, N. P., & War, N., "Text Line Segmentation on Myanmar Handwritten Documents Using Directional Gaussian Filter," 2024 IEEE Conference on Computer Applications (ICCA), IEEE, 2024.
[18] Xie, Z. F., Lau, R. W., Gui, Y., Chen, M. G., & Ma, L. Z., "A gradient-domain-based edge-preserving sharpen filter," The Visual Computer, vol. 28 pp. 1195-1207, 2012.
[19] Verma, A., Dhanda, N., & Yadav, V., "Binary particle swarm optimization based edge detection under weighted image sharpening filter," International Journal of Information Technology, vol. 15, no. 1, pp. 289-299, 2023.
[20] Ranjan, R., & Avasthi, V., "Edge detection using guided Sobel image filtering," Wireless Personal Communications, vol. 132, no. 1, pp. 651-67, 2023.
[21] Ramponi, G., Strobel, N. K., Mitra, S. K., & Yu, T. H., "Nonlinear unsharp masking methods for image contrast enhancement," Journal of electronic imaging, vol. 5, no. 3, pp. 353-366, 1996.
[22] Kadomoto, S., Nanegrungsunk, O., Nittala, M. G., Karamat, A., & Sadda, S. R., "Enhanced detection of reticular pseudodrusen on color fundus photos by image embossing," Current eye research, vol. 47, no. 11, pp. 1547-1552, 2022.
[23] Soni, A., & Rai, A., "Kidney stone detection and extraction using directional emboss & SVM from computed tomography images," 2020 Third International Conference on Multimedia Processing, Communication & Information Technology (MPCIT), IEEE, 2020.
[24] Tsai, D. M., & Huang, C. K., "Defect detection in electronic surfaces using template-based Fourier image reconstruction," IEEE Transactions on Components, Packaging and Manufacturing Technology, vol. 9, no. 1, pp. 163-172, 2018.
[25] Kaur, K., Jindal, N., & Singh, K., "Fractional Fourier Transform based Riesz fractional derivative approach for edge detection and its application in image enhancement," Signal Processing, vol. 180, pp. 107852, 2021.
[26] Petersen, A., Gelb, A., & Eubank, R., "Hypothesis testing for Fourier based edge detection methods," Journal of Scientific Computing, vol, 51, pp. 608-630, 2012.
[27] Canny, J., "A computational approach to edge detection," IEEE Transactions on pattern analysis and machine intelligence, vol. 6, pp. 679-698, 1986.
[28] Xu, Q., Varadarajan, S., Chakrabarti, C., & Karam, L. J., "A distributed canny edge detector: algorithm and FPGA implementation," IEEE Transactions on Image Processing, vol. 23, no. 7, pp. 2944-2960, 2014.
[29] Kim, Y. W., & Krishna, A. V., "A study on the effect of Canny edge detection on downscaled images," Pattern Detection and Image Analysis, vol. 30, pp. 372-381, 2020.
[30] Sekehravani, E. A., Babulak, E., & Masoodi, M., "Implementing canny edge detection algorithm for noisy image," Bulletin of Electrical Engineering and Informatics, vol. 9, no. 4, pp. 1404-1410, 2020.
[31] Hough, P. V., "Method and means for detecting complex patterns," U.S. Patent, No. 3,069,654, 18 Dec, 1962.

[32] Marzougui, M., Alasiry, A., Kortli, Y., & Baili, J., "A lane tracking method based on progressive probabilistic Hough transform," IEEE access, vol. 8, pp. 84893-84905, 2020.
[33] Zhao, K., Han, Q., Zhang, C. B., Xu, J., & Cheng, M. M., "Deep hough transform for semantic line detection," IEEE Transactions on Pattern Analysis and Machine Intelligence, vol. 44, no.9, pp. 4793-4806, 2021.
[34] Lin, Y., Pintea, S. L., & Van Gemert, J. C., "Deep hough-transform line priors," Computer Vision–ECCV 2020, 16th European Conference, Glasgow, UK, August 23–28, 2020, Proceedings, Part XXII 16, Springer International Publishing, 2020.
[35] Dahyot, R., "Statistical hough transform," IEEE Transactions on pattern analysis and machine intelligence, vol. 31, no.8, pp. 1502-1509, 2008.
[36] T.S. Arulananth, P. Chinnasamy, J. Chinna Babu, Ajmeera Kiran, J. Hemalatha, & Mohamed Abbas, "Edge detection using fast pixel based matching and contours mapping algorithms," PLoS ONE, vol. 18, no.8, 2023.
[37] Yin, Z., Wang, Z., Fan, C., Wang, X., & Qiu, T., "Edge Detection via Fusion Difference Convolution," Sensors, vol. 23, no.15, 2023.